\documentclass[times,11pt,psfig,twoside]{acta}
\usepackage{amsmath}
\usepackage{euscript}
\usepackage{graphicx} 
\usepackage{pslatex}
\SetPages{345}{358}
\SetVol{58}{2008}
\begin{document}

\begin{Titlepage}

\centerline{\bf Binary Source Lensing and the Repeating OGLE EWS Events}

\Author{M.~~ J~a~r~o~s~z~y~\'n~s~k~i ~~ and ~~J.~~ S~k~o~w~r~o~n
}
{
Warsaw University Observatory, Al.~Ujazdowskie~4,~00-478~Warszawa, Poland\\
e-mail: (mj,jskowron)@astrouw.edu.pl
}

\Received{December 19, 2008}

\end{Titlepage}

\Abstract{A microlensing event may exhibit a second brightening when the
source and/or the lens is a binary star. Previous study revealed 19 such
repeating event candidates among 4120 investigated microlensing light 
curves of the Optical Gravitational Lensing Experiment (OGLE).
The same study gave the probability $\approx 0.0027$ for a repeating
event caused by a binary lens. We present the simulations of binary
source lensing events and calculate the probability of observing a
second brightening in the light curve. Applying to simulated light
curves the same algorithm as was used in the analysis of real OGLE data,
we find the probability $\approx 0.0018$ of observing a second
brightening in a binary source lensing curve. The expected and measured
numbers of repeating events are in agreement only if one postulates that
all lenses and all sources are binary. Since the fraction of binaries is
believed to be $\le 50$\%, there seems to be a discrepancy. 
}{Gravitational lensing - Galaxy structure - binaries: general}

\Section{Introduction}

Investigations of the Galactic disk show that 
the fraction of stars in binary systems is high, reaching 57\% for
solar-type stars (Duquennoy and Major, 1991). For later spectral types
the multiplicity fractions seem to be lower (Fisher and Marcy 1992; Reid
and Gizis 1997). Lada (2006) summarizes the existing observations,
concluding that two thirds of the main sequence stars in the disk have
no companions.

The shape of the microlensing light curve may be changed if the observed
source is a binary star. In particular the event may look as subsequent
two events corresponding to microlensing of two binary components.
The probability of observing a binary source microlensing has been investigated
theoretically by several authors (Griest and Hu, 1992; Dominik, 1998; Han
and Jeong, 1998; Han 2005, to cite few). The typical conclusion of such
studies is that in few percent of cases where the source is a binary, 
this fact should have observable consequences. Assuming that the stars
in the Galactic bulge form binary systems with probability similar to
the stars in the disk (which has not been proved), one expects that
several tens of events among several thousands discovered to date have
binary source characteristics.

In a recent paper Skowron \etal (2009) investigate the problem of
repeating microlensing events  using the data of the Optical Gravitational
Lens Experiment (\eg Udalski 2003). Majority of the events have been
discovered by the Early Warning System (EWS - see Udalski \etal 1994;
Udalski 2003 for details). 
The definition of an {\it repeating microlensing event}
adopted by Skowron \etal requires that the two brightenings in the
light curve caused by microlensing are well separated, i.e. the
observed luminosity comes to the baseline between the peaks. There are
19 such cases among 4120 events investigated, and 12 of them are
interpreted as a result of lensing by wide binary systems, while 6 allow
concurrent binary lens/binary source interpretations. Even if all
ambiguous cases were binary source events, the probability of observing
a binary source repeating event would be only 0.15\%.

Not all the binary source events comply with the above definition of
repeating events. If the binary source separation is of the order of the
lens Einstein radius, the peaks in the light curve partially overlap.
This may produce smooth light curves of various shapes, some
indistinguishable from those produced by single source approaching cusps
of the binary lens caustics. Even the binary lens caustic crossing event
may be classified as due to binary source if the observations are
sparse. Examples of several binary source / binary lens events are given by
Jaroszyñski \etal (2004, 2006) and Skowron \etal (2007).

In this paper we simulate the microlensing light curves with binary
sources to find probability of producing a repeating event. Our aim is
to obtain realistic light curves, and we include several details typical
for the OGLE team observations. (Sampling rates, observational errors
and their dependence on observed luminosity, span of observations etc).
Also the classification of simulated light curves and fitting
procedures closely follow the approach of Skowron \etal (2009). We
expect, that the probability of finding a repeating event among
simulated light curves would be directly comparable with the number
obtained as a result of analyzing all EWS events by Skowron \etal
(2009), and some limits on the binary stars population in the Galactic
bulge will be possible to place.

\Section{The simulations}

\subsection{Parametrization of the light curves}

A binary source event light curve can be modeled using eight parameters:
the times of closest approaches of the lens to the source components
$t_{01}$ and $t_{02}$, the characteristic Einstein time $t_\mathrm{E}$,
the two dimensionless impact parameters $b_1$ and $b_2$, the two source
components energy fluxes $F_1$ and $F_2$, and the blended flux of stars
in the seeing disk $F_\mathrm{b}$. The observed flux is a given function
of time:
\begin{equation}
F(t)=A(u_1(t))F_1 + A(u_2(t))F_2 + F_\mathrm{b}
\end{equation}
where $A(u)$ describes the so called Paczyñski curve (e.g. Paczyñski, 1991):
\begin{equation}
A(u)=\frac{u^2+2}{u\sqrt{u^2+4}}~~~~~~
u_i(t)=\sqrt{\left(\frac{t-t_{0i}}{t_\mathrm{E}}\right)^2+b_i^2}
\end{equation}
The model parameters listed above are related to even larger number of
parameters characterizing the lens and source distances $d_\mathrm{L}$,
$d_\mathrm{S}$, lens mass $m$, observer, lens and source velocities 
$\mathbf{v}_\mathrm{O}$, $\mathbf{v}_\mathrm{L}$,
$\mathbf{v}_\mathrm{S}$, and source components intrinsic luminosities
$L_1$, $L_2$. The Einstein time is is given as
\begin{equation}
t_\mathrm{E}=\frac{r_\mathrm{E}}{|\mathbf{v}_\perp|}
~~~~~~~~~r_\mathrm{E}=\sqrt{\frac{4Gm}{c^2}
\frac{d_\mathrm{L}(d_\mathrm{S}-d_\mathrm{L})}{d_\mathrm{S}}}
\end{equation}
where $r_\mathrm{E}$ is the Einstein radius and $\mathbf{v}_\perp$ is
the velocity of the lens relative to the line joining the observer and the
source, measured in the plane perpendicular to this line:
\begin{equation}
\mathbf{v}_\perp
=\left(\mathbf{v}_\mathrm{L}-(1-x)\mathbf{v}_\mathrm{O}
-x\mathbf{v}_\mathrm{S}\right)_\perp
~~~~~~~~~x\equiv \frac{d_\mathrm{L}}{d_\mathrm{S}}
\end{equation}
The Einstein radius is a unit of length in the lens plane; in the source
plane the projected Einstein radius ${\tilde r}_\mathrm{E} \equiv
r_\mathrm{E}/x$ plays the same role. 

The shape of the light curve depends on the binary source separation or,
more precisely, on the projected into the sky distance between the two
components in the time of interest, expressed in Einstein radius units,
$d \equiv a_\perp/{\tilde r}_\mathrm{E}$. Since we investigate a
population of binary sources with large scatter of separations and
the microlensing events may happen in any phase of binary motion,
we replace the separation by the binary semimajor axis.

The angle between the source trajectory and the line joining its
components,  both projected into the sky, has a random value 
$0 \le \beta \le 2\pi$. Once it is chosen, the separation perpendicular
to the trajectory $d_\perp \equiv |d\sin\beta|$ can be calculated. 

Since we are interested in observed fraction of binary source microlensing
events among events which may be interpreted as due to a single source, we
assume that the lens passes within projected Einstein radius from
at least one of the binary components. Depending on the value of
$d_\perp$, the region of interest may consist of one ($d_\perp<2$) or
two parts ($d_\perp>2$). We assume that the possible trajectories cross
any point of the region of interest with the same probability. The
choices of $d$ and $\beta$ imply the following relations between the
parameters: 
\begin{equation}
b_2=b_1 + d\sin\beta~~~~~~~~~~
t_{02}=t_{01}+d~t_\mathrm{E}\cos\beta
\end{equation}
and (by construction) at least one of the inequalities $|b_1|\le 1$ 
or $|b_2|\le 1$ holds. We choose $t_{01}$ at random from the time 
interval during which the observations were
performed, so $t_\mathrm{start}\le t_{01} \le t_\mathrm{end}$, where
$t_\mathrm{start}$, $t_\mathrm{end}$, denote the beginning and end of
observations. 

\subsection{Distributions of physical parameters}

Among the physical parameters, only the Einstein time $t_\mathrm{E}$, 
the dimensionless binary separation $d$ and the fluxes $F_1$, $F_2$, 
and $F_\mathrm{b}$ have a direct influence on the shape of the observed
(simulated) light curve and may be correlated with each other. The other
light curve parameters depend on random choices of values and are
independent of the parameters listed above.

We find the distribution of the Einstein time-scales using the
simulations  of the Galaxy structure (see below).  The influence of the
extinction and its dependence on position makes it difficult to obtain
the distribution of apparent luminosities for the OGLE sources from a
simple Galaxy model, and we use the distribution of baseline fluxes from
the OGLE observations. The distribution of binary sources separations 
must be found independently, based on the model of binary systems
population. Since the dimensionless separation $d$ depends also on the
value of the Einstein radius, the simulations of the Galaxy structure
are needed again. We also find the distribution of distance moduli to
sources, to check their correlations with other parameters.

The population of Galaxy disk binary stars has been studied by many
authors (Abt 1983, Duquennoy and Mayor 1991, Fisher and Marcy 1992
Lada 2006 and references therein).
We adopt a log-normal distribution of binary component
separation $a$ with $<\log a> \approx 1.5$ and $\sigma_{\log a} \approx 1.5$
after Han and Jeong (1998), who use the study of Duquennoy and Mayor
(1991). Little is known about the binaries in the bulge, and we assume
that their population is similar to the disk one.

We use the Han and Gould (2003) model of the density distribution in the
Galaxy disk. We also adopt their mass function (the same for the disk
and the bulge), which includes brown
dwarfs and main sequence stars with continuous distributions of masses,
as well as discrete contributions from black holes ($5M_\odot$), neutron
stars ($1.35M_\odot$) and white dwarfs ($0.6M_\odot$). For the bulge
density distribution we use the triaxial exponential model E2 of 
Dwek \etal (1995) with parameters given by Stanek \etal (1997).

We basically follow Wood and Mao (2005) description of the disk and bulge
kinematics. The adopted parameters are: disk rotation speed
$v_\phi=220\mathrm{km/s}$, with the dispersion
$\sigma_\phi=30\mathrm{km/s}$, $\sigma_z=20\mathrm{km/s}$ in the 
azimuthal and perpendicular to the disk plane directions respectively.
Inside the central $0.5\mathrm{kpc}$ the flat rotation curve is
replaced by the solid body rotation curve. For the bulge we assume solid
body rotation up to $0.5\mathrm{kpc}$, followed by flat rotation curve
with velocity $v_\phi=50\mathrm{km/s}$ at greater radii. The
velocity dispersion for the bulge is:
$\sigma_{x,y,z}=(110, 82.5, 66.3)\mathrm{km/s}$. Both the disk and the
bulge kinematics descriptions are simplified (compare e.g. Sofue, Honma
and Omodaka 2008 for the disk, and Clarkson \etal 2008 for the bulge),
but, as our numerical experiments show, the details have little
influence on the obtained distributions of the Einstein times and radii.

We use Monte Carlo method to get the relations between the intrinsic
physical parameters influencing the microlensing phenomena. Since the
observed distribution of parameters depends on their probability of
occurrence, we always weight any set of intrinsic parameters by their
contribution to the rate of events (Griest 1991, Kiraga and Paczyñski 
1994, Wood and Mao 2005):
\begin{equation}
\Delta\Gamma \propto n_\mathrm{L}(m)\Delta m \Delta d_\mathrm{L}
~n_\mathrm{S}d_\mathrm{S}^2 \Delta d_\mathrm{S}~v_\perp~r_\mathrm{E}
\end{equation}
where $n_\mathrm{L}(m)$ is the number density of lenses with masses in
the range $m$ --- $m+\Delta m$, at the distance $d_\mathrm{L}$ from the
observer, $n_\mathrm{S}$ is the number density of "observable" sources
at the distance $d_\mathrm{S}$, $v_\perp$ is the lens velocity, and
$r_\mathrm{E}$ - the Einstein radius. The number density of observable
sources is proportional to mass density times the fraction of all stars,
which are brighter than
$M_I=I_\mathrm{lim}-5\lg(d_\mathrm{S}/10\mathrm{pc})$, where we use
$I_\mathrm{lim}=20$ as a faint limit of OGLE photometry. We use Holtzman
\etal (1998) luminosity function for the bulge and Gould, Bahcall, and
Flynn (1997) for the disk.

We investigate the vicinity of the Galaxy center choosing at random
locations with Galactic longitudes $-8^\circ \le l \le 10^\circ$ and
latitudes $-7.5^\circ \le b \le -1.5^\circ$, which includes (but is not
identical to) the majority OGLE fields. The source distance is also 
chosen at random
from the range $1 \le d_\mathrm{S} \le 13~\mathrm{kpc}$, and the lens
has a random location between the observer and the source. We assume
that the probability that an object at given position belongs to the
bulge or to the disk population is proportional to the respective mass
densities, and each choice is done with random number generator. 
Thus four different combinations (source in the bulge and lens
in the bulge i.e. "bulge - bulge", and similarly "bulge - disk", "disk -
bulge", and "disk - disk") are possible. The velocities are assigned
depending on the population to which the source/lens belong.
The Gaussian part of each velocity component is chosen using
Gaussian number generator. The lens mass is assigned using reversed
cumulative mass distribution. The double source separation has
log-normal distribution, so it also can be obtained with Gaussian number
generator. We use random number and Gaussian generators as advocated by
Press \etal (2007).

The choices of intrinsic parameters described above define the
values of parameters closely related to observations, 
such as the Einstein time $t_\mathrm{E}$, the dimensionless source
separation $d$, or the source distance modulus $m-M \equiv
5\lg(d_\mathrm{S}/10\mathrm{pc})$. We construct three-dimensional
histograms for the expected occurrence of these parameters, 
storing independently information for the four combinations of the 
source/lens belonging to the bulge/disk populations. 

\begin{figure}[htb]
\begin{center}
\includegraphics[width=91.0mm]{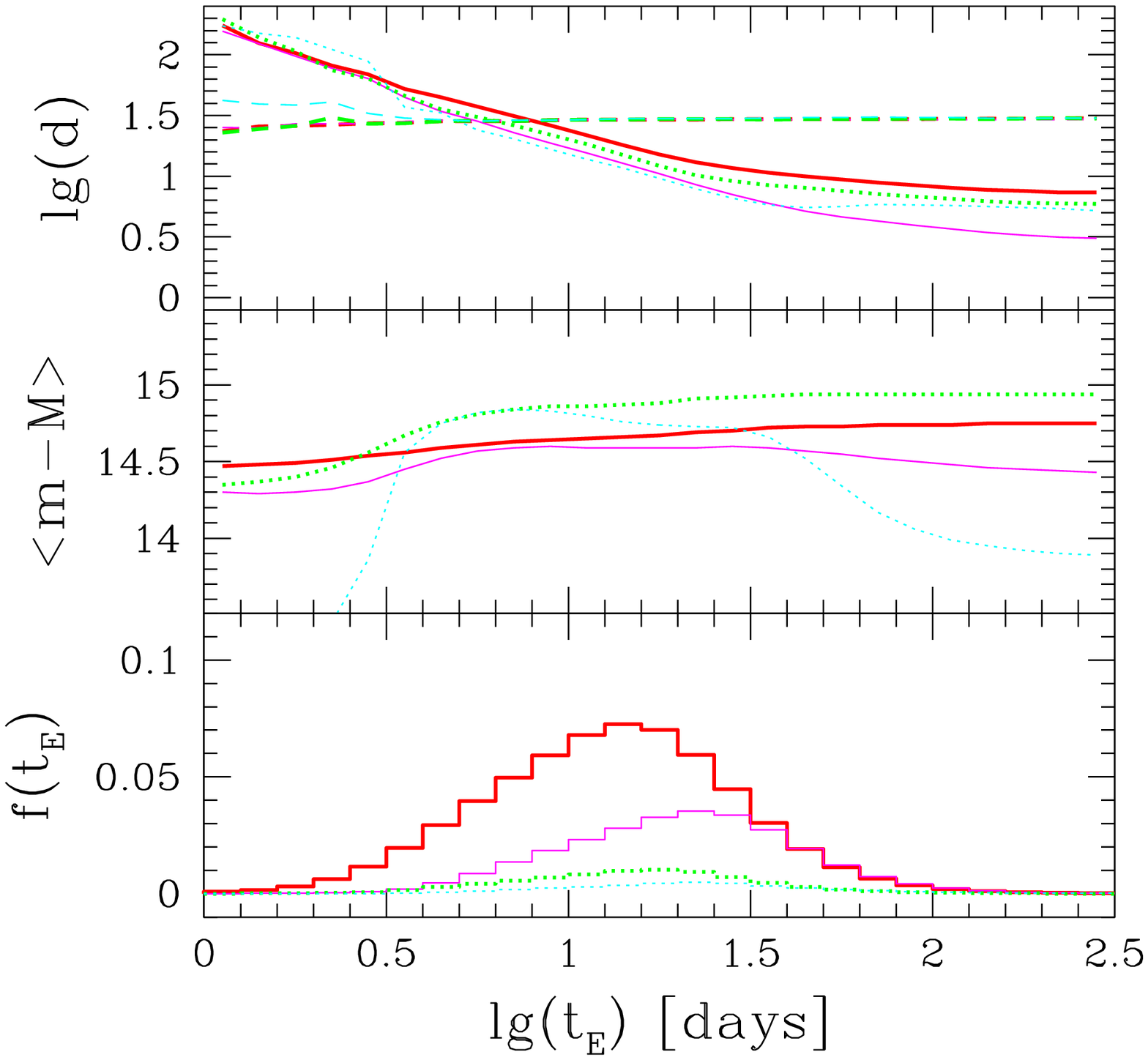}
\end{center}
\FigCap{The dependence of averaged source separation $\lg(d)$ (upper
panel) and distance modulus (middle) on the Einstein time logarithm. 
The bottom panel shows histogram of Einstein time logarithm. In each of
the panels four cases are considered: "b-b" (thick solid lines), "b-d"
(light solid), "d-b" (heavy dotted), and "d-d" (light dotted). (See the
text for naming convention.) In the upper panel also the dispersions 
of source separations are drawn with dashed lines.}
\end{figure}

The results are shown in Fig.1. We treat the Einstein time,
which can be directly measured in the model fits to light
curves, as a primary variable. The histograms in the bottom panel show
its distribution for the four cases mentioned. The relative importance
of the four cases is given as the ratio: b-b:b-d:d-b:d-d=57:27:11:5,
where "b" stays for the bulge and "d" - for the disk. 
The resulting distribution of the Einstein times logarithm is well
approximated by a Gaussian with the average $<lg(t_\mathrm{E})>=1.16$ 
and the dispersion $\sigma(\lg(t_\mathrm{E}))=0.30$. This is the
simulated (in a sense {\it intrinsic}) distribution of Einstein times. 

In the upper panel the averaged dimensionless source separation is shown
as a function of the Einstein time. The dependence is strong and
slightly different for the "cases" considered. On the other hand the 
distribution of separations in each small $\Delta t_\mathrm{E}$ interval
remains log-normal, and the dispersion is practically independent of the
time scale or case considered.

The middle panel shows that the distance modulus to the source is only
weakly correlated with the Einstein timescale, except the least
important case of the source and the lens both belonging to the disk.
This can be understood as a direct consequence of the limited spatial
extent of the bulge.

The relations found above suggest, that it is safe to simulate binary
source light curves ignoring the correlations of observed luminosity
with the timescale, but it is necessary to take into account the
dependence of the source components separation on this parameter.

Assuming the similarity of the bulge binaries to better known binaries
from Solar neighborhood, we postulate the mass luminosity relation for
component stars in the form $L \sim m^\alpha$, which implies that the
luminosity ratio, $w \equiv L_2/L_1$ is the function of the mass ratio 
$q\equiv m_2/m_1$: $w=q^\alpha$. Using stellar $I$ - band magnitudes 
one has:
\begin{equation}
\Delta I \equiv I_2-I_1 = 2.5\alpha\lg q
\end{equation}
Assuming that the logarithm of the mass ratio is uniformly distributed
in the range corresponding to $0.1 \le q \le 1$, (compare Skowron \etal,
2009) and postulating that $\alpha \approx 3.2$, we
see that magnitude difference is uniformly distributed in the
range $0 \le \Delta I \le 8$. We use this distribution of luminosity
ratios in our simulations.

The definition of the blended flux in the case of a binary source is not
obvious. For the binary sources undergoing microlensing and interpreted
as such the definition of blending parameter is natural:
\begin{equation}
f \equiv \frac {F_1+F_2}{F_1+F_2+F_\mathrm{b}}
\end{equation}
where the values of fluxes are obtained from the fit.
If the same source undergoes microlensing interpreted as single source
event involving its first component, the fitted blending parameter
would be:
\begin{equation}
f \equiv \frac {F_1}{F_1+F_2+F_\mathrm{b}}
\end{equation}
and $F_2+F_\mathrm{b}$ would serve as blended flux (compare Dominik,
1998). Thus it is not clear which blending parameter should be used and
what distribution for its values should be adopted. For the sake of
simplicity we use the first of the definitions, thus avoiding possible
dependencies between components flux ratio and blending parameter. 
We also assume the blending parameter to be uniformly distributed, as
suggested by fits to a limited sample of binary lens events
(Jaroszyñski \etal 2004).

We get the baseline flux $F_\mathrm{base}$ from the condition 
$P(>F_\mathrm{base})=\mathcal{R}$, where $\mathcal{R}$ symbolizes the
random number generator and a result of its call, and
$P()$ denotes the cumulative probability distribution valid for all OGLE
EWS events. Next the blending parameter $f=\mathcal{R}$ is obtained with
the subsequent random number. 

In principle the components stellar magnitude difference could also be
chosen at random, since we have assumed its uniform distribution, but
it is less time consuming to consider a finite number of fixed flux
ratios, and derive the probability of discovering a repeating event for
each of them. Binaries of similar luminosity components produce binary
source events with higher probability (Han and Jeong, 1998), as compared
to binaries with large flux ratios, so they require smaller number of
simulations for the same accuracy of the result.

The fluxes relevant for the light curve simulation are given as:
\begin{equation}
F_1=\frac{1}{1+w}fF_\mathrm{~base}~~~
F_2=\frac{w}{1+w}fF_\mathrm{~base}~~~
F_\mathrm{b}=(1-f)F_\mathrm{~base}
\end{equation}
where $w$ is one of the considered values of the flux ratio.

\subsection{Light curves}

Once all relevant parameters are chosen the model light curve of a
binary source lensing event can be obtained using Eqs. 1 and 2. To
simulate the observations we~use a reference light curve from the OGLE
EWS database.  (We have employed several reference light curves for
several independent series of simulations.) 
The reference light curve gives the observed luminosities
$I_\mathrm{ref}$ and estimated errors $\Delta I_\mathrm{ref}$ at a
series of observation times $t_i$. Using our equations we can get the
model luminosities $I_\mathrm{mod}$ for each $t_i$. The simulated
observational errors are obtained by rescaling the reference errors
according to the formula of Wyrzykowski (2005):
\begin{equation}
\Delta I_\mathrm{mod}=\Delta I_\mathrm{ref}
10^{0.33875(I_\mathrm{mod}-I_\mathrm{ref})}
\end{equation}
which describes the dependence of the estimated observational errors on
measured luminosities.

The comparison between the fluctuations in luminosity of a set of
constant flux stars and estimated measurement errors given by the same
OGLE photometry pipeline (Wyrzykowski \etal 2008) suggests a further
rescaling of the errors. The OGLE errors are slightly underestimated and
we adopt:
\begin{equation}
\Delta I =\sqrt{(1.38 \Delta I_\mathrm{ref})^2+0.0052^2}
\end{equation}
as a measure of observational errors, following Skowron \etal (2009).

Finally we add a Gaussian noise to the model light curve:
\begin{equation}
I=I_\mathrm{mod}+\Delta I~\mathcal{G}
\end{equation}
where $\mathcal{G}$ is a Gaussian number generator with zero mean and unit
dispersion. The set of $I(t_i)$ and $\Delta I(t_i)$ represents a
simulated light curve. We use its flux equivalent for further analysis.

\subsection{Looking for repeating microlensing events}

The strategy used here closely follows the method of analysis applied to
the real OGLE EWS light curves by Skowron \etal (2009) in search for
repeating microlensing events. The search has been done with the help of
an automated algorithm which is also employed here. Since the
non-microlensing light curves are absent among our simulated set, the
visual inspection of the events picked up by the algorithm does not seem
to be necessary.

First we fit a constant flux model to the simulated light curve and
calculate its $\chi^2(\mathrm{constant})=\chi^2_0$, which serves as
reference for other models. Next we find the main peak in the light
curve and try a single source fit, obtaining
$\chi^2(\mathrm{single})=\chi^2_1$. Our previous experience shows that
for $\sim 10^3$ observations, the attempt of fitting a single source model to
the constant flux + noise light curve typically improves the fit by
$\Delta\chi^2 \approx 15$, and $\Delta\chi^2 > 55$ happens in only 1\%
of cases. Thus if $\chi^2_0 - \chi^2_1 < 55$ we conclude that the
synthetic light curve does not represent a microlensing event, which may
be caused by excessive blending, passing close to the very faint binary
component, or due to the fact that the event takes place out of
observational season.

If the single source model improves the fit sufficiently, we try to
remove the main peak to look for another one. To remove the main peak we
use the single source model and find the time span where model light
curve is more than $3\sigma$ above the baseline flux level. All
simulated fluxes from this time interval are replaced by the baseline
flux plus Gaussian noise, with observational errors rescaled to the base
level. A constant flux model is again fitted to the modified light
curve for comparison with the single source fit applied next. Since we
are going to compare simulation results with the Skowron \etal (2009)
analysis of real microlensing events, we follow their approach in
identifying and evaluating the second peak. We require the following
inequality to hold:
\begin{equation}
\chi^2_0-\chi^2_1 \ge 0.2\chi^2_1
\end{equation} 
where $\chi^2_0$ and $\chi^2_1$ are related to the constant flux and
single source models of the modified light curve respectively. This
(rather strong) requirement produces samples of repeating microlensing
events similar to obtainable by visual light curve inspection, as shown
by Skowron \etal (2009).

If the improvement of the fit does not meet our requirement, we
conclude that the second peak is absent, and that we probably deal with 
a single source event. If the original single source fit for the main peak is
satisfactory ($\chi^2/N_\mathrm{DOF}$ not too high) we write down its
parameters as well as the original simulation parameters for future analysis.
Otherwise the case is rejected.

On the other hand, if the existence of the second peak is established 
according to our criterion, we 
fit a double source model to the original light curve. We use the
positions of the peaks, their impact parameters, and the fluxes obtained in the
single source fits as starting parameter values for the final model. The
starting value for the Einstein time is taken from the single source fit
to the main peak, and the blended flux is estimated from the flux
budget ($F_1+F_2+F_\mathrm{b} \equiv F_\mathrm{base}$ for a correct model).
The goodness of fit for the binary source is checked. Finally we check
whether the double source model light curve comes to the baseline,
finding its minimum flux $F_\mathrm{min}$ between the peaks and checking
whether $F_\mathrm{min} \le F_\mathrm{base}+3\sigma_\mathrm{base}$. If
the case passes both tests we treat it as a repeating double source
event. If goodness of fit test is passed but baseline test is not, the
case is treated as non-repeating double source event. In any case the
parameters of the fits and original simulation parameters are written
down for further analysis. It must be admitted that our method is
primarily looking for repeating events and many double source events are
lost by our algorithm. In particular, if the peaks are partially
overlapping, they would probably be both fitted by a single source model 
and the second peak would be lost. In consequence the case may be
classified as a single source event or rejected due to the low fit quality. 

\section{Results}

Using the described methods we have simulated few millions of binary source
lensing light curves. About one third to one half of the simulated cases
(depending on the flux ratio)  have been classified by
our algorithm as microlensing event candidates, while the other have
been rejected because of too weak or missing microlensing signatures.
This may have few reasons. We simulate binary source events and in half
of the cases the lens passes close to the fainter component, thus
leading to possibly insignificant magnification of the total flux. The
added blended flux may have similar effect. The simulated magnification
may happen off season, also preventing identification.

The distribution of the Einstein times used in simulations has been
obtained from the Galaxy models. We always use distributions of
time-scale logarithms in our calculations, and the mean values and
standard deviations reported below are always mean logarithms and their
deviations. (See also Discussion.)

Our Galaxy model gives the intrinsic distribution of
Einstein times for all would-be lenses with mean logarithm corresponding
to $\approx 14^\mathrm{d}$. The subsample of microlensing candidates has
the average corresponding to $18^\mathrm{d}$, where the simulation
parameter values are used. Finally, the average for parameters fitted to
the same subsample corresponds to $22^\mathrm{d}$. The three
distributions are shown in Fig.2. 
The standard deviations for two model
distributions are the same (0.3 in logarithm corresponding to factor 2
in Einstein times) and for the distribution of fitted parameters
slightly higher (0.5 or factor 3). This systematic shift to longer
time-scales may be understood as the effect of preferential rejection of
short lasting events, since they are easier to miss, and as a result of
Wo¼niak and Paczyñski (1997) degeneracy, which allows longer lasting,
more blended and more strongly amplified models of the same fit
quality.
The small subsample of binary source lensing repeating events has still
longer time-scales, with average corresponding to $26^\mathrm{d}$, without
significant difference between simulations and fits, and with deviation
$\sigma_{\log} =0.3$.

The effectiveness of finding a repeating double source event as
a function of the binary components stellar magnitude difference is shown
in Fig.3. It is a sharply decreasing function of luminosity difference
and systems with $\Delta I >4^\mathrm{m}$ have negligible chance of
producing a repeating event. This tendency is in agreement with the
results of Han and Jeong (1998). 

The main result of our study is the evaluated probability of finding a
repeating microlensing event among large number of events caused by
sources belonging to the population of binaries of assumed properties. 
The effectiveness averaged over the interval of eight stellar magnitudes
is 0.18\% and this number should be compared with the result obtained for
the sample of OGLE EWS events by Skowron \etal (2009). 

\begin{figure}[htb]
\begin{center}
\includegraphics[width=83.0mm]{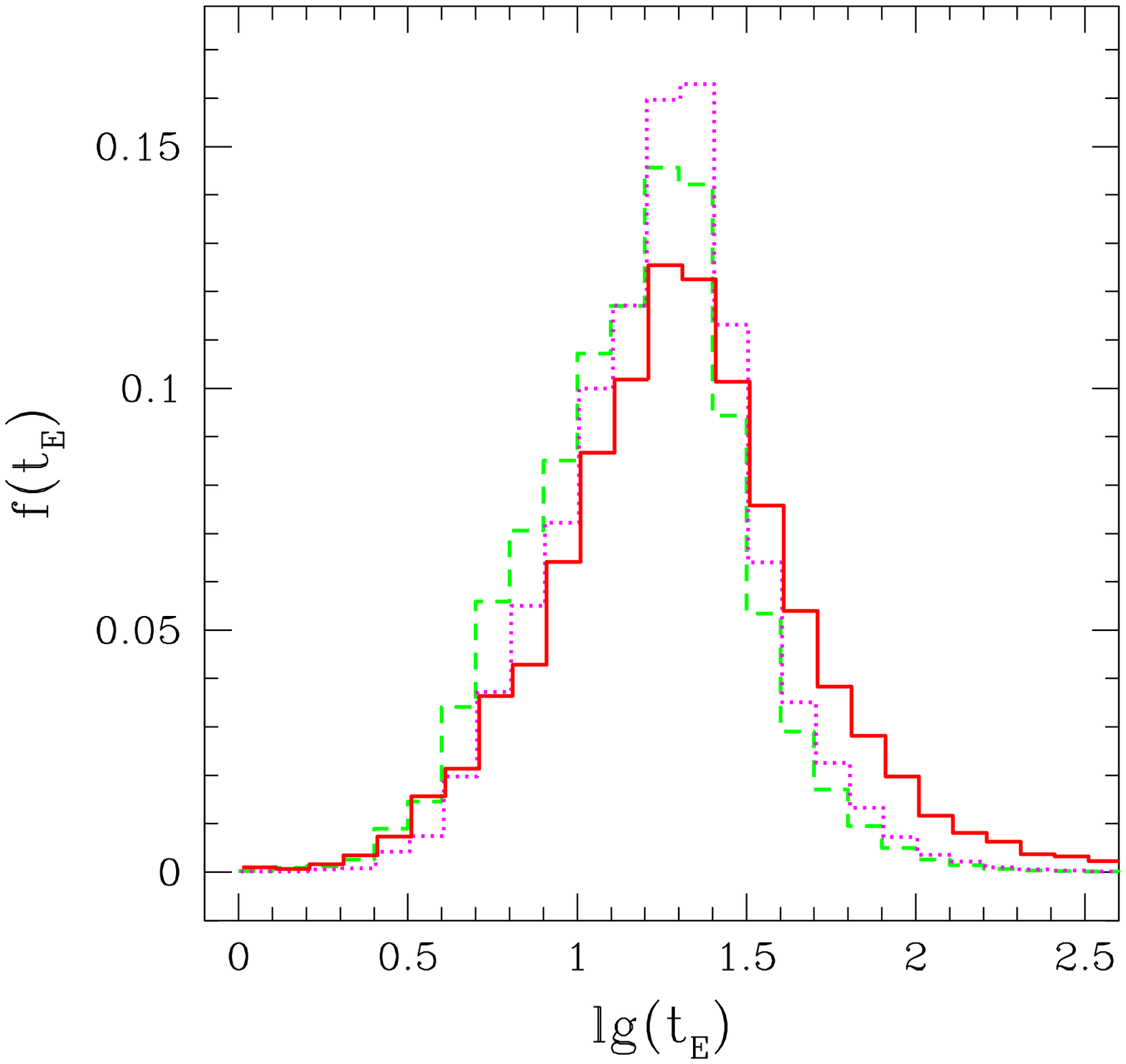}
\end{center}
\FigCap{The histograms of the Einstein time. The dashed line represents
the distribution for all simulated models. The dotted line also shows the
distribution of model parameter values, but is limited to cases accepted
as microlensing events. The solid line shows the distribution of fitted
parameter values, again limited to accepted models.}
\end{figure}

\begin{figure}[hbt]
\begin{center}
\includegraphics[width=83.0mm]{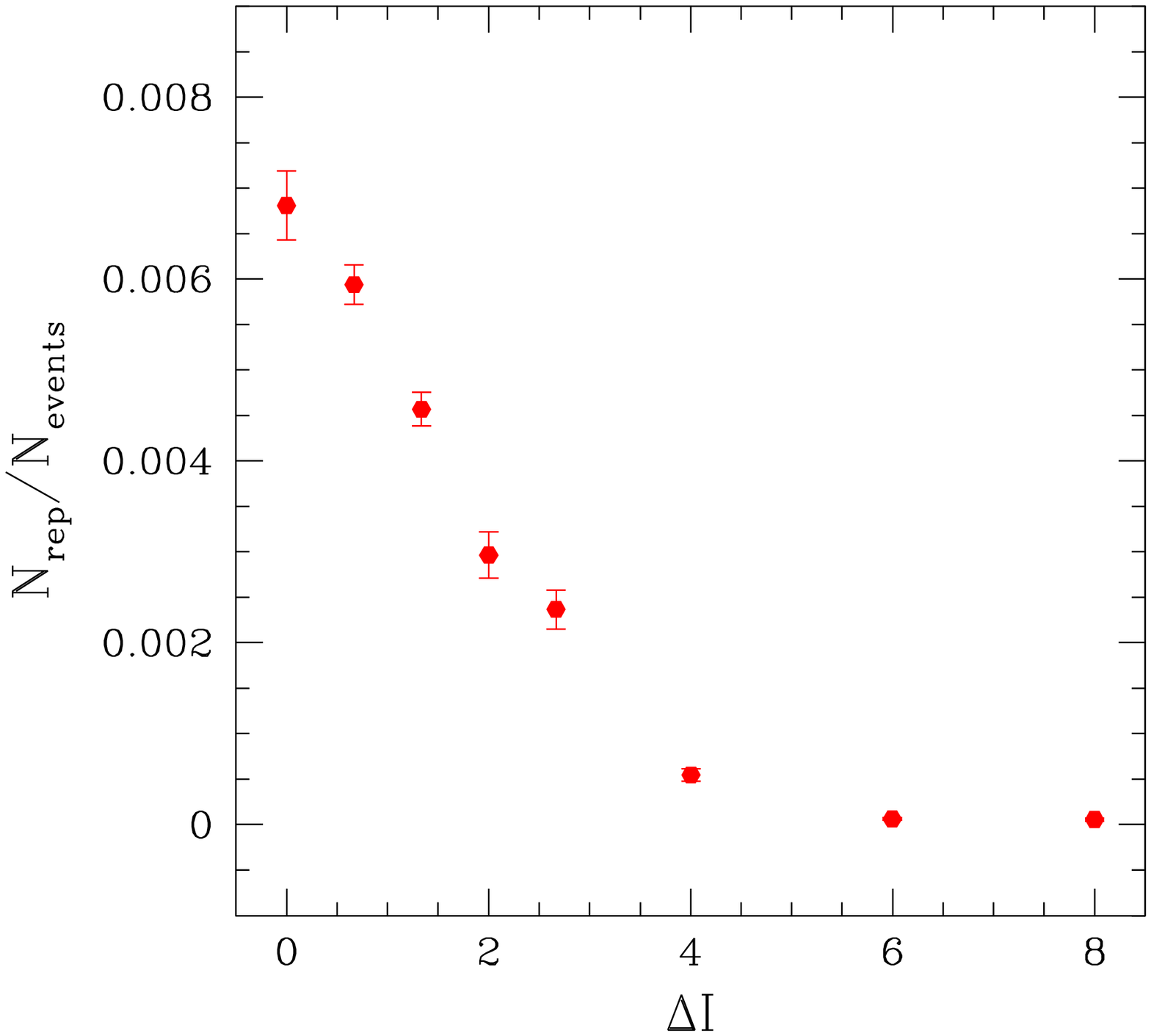}
\end{center}
\FigCap{The effectiveness of discovering a repeating microlensing event
as a function of the binary source components flux ratio. The flux ratio
is expressed in stellar magnitudes. ($\Delta I \equiv |2.5\lg(F_1/F_2)|$).
The error bars are estimated based on the Poisson statistics and the
number of repeating events found for given flux ratio.}
\end{figure}


\section{Discussion}

We have calculated the probability of observing a {\it repeating}
microlensing event among the sample of OGLE EWS events toward the Bulge,
caused by the binary nature of the source star. If all the sources
included in the sample were binary, there should be $4120*0.0018 \approx
7$ repeating microlensing events of this kind, where the estimate is
done for the 4120 events investigated by Skowron \etal (2009). 
(The events of season 2008 have been excluded from their study.)

Analogous calculation by Skowron \etal (2009) shows that if all the
lenses causing OGLE sample events were binary, they should produce
another $4120*0.0027 \approx 11$ repeating events. The analysis of
light curves from the sample by the same authors gives 19 repeating
events, 12 probably due to binary lenses, 6 possibly due to binary
sources, and one leading to unsatisfactory models.

Since not all stars belong to multiple systems, there seems to be a
problem with the interpretation of the repeating events. If half of
the systems were binary, both in the disk and in the bulge (which is    
probably an overestimation -- compare Introduction),  the expected
numbers of repeating events would be roughly 6 plus 3 due to binary
lenses and binary sources respectively. According to Poisson statistics
probability of observing 18 cases instead of expected 9 is only
$0.0029$. 
 
Our result is valid as long as our assumptions are true. The strong
dependence of the probability of observing a repeating event on the
binary source flux ratio, as well as a similar dependence of the
probability on the binary lens mass ratio obtained by Skowron \etal 
(2009) suggests, that one may avoid the discrepancy narrowing the ranges
of the flux / mass ratios. For example assuming that population of
binary sources is limited to $0 \le \Delta I \le 4$ with uniform
distribution increases the averaged effectiveness twofold, since extreme
flux ratio binaries with $4 \le \Delta I \le 8$ can produce only negligible
number of repeating events. (Similarly narrowing the range of binary
lenses mass ratios increases their probability of producing repeating
events).  The simple narrowing of the parameter ranges is only an
example, but it shows that a different distributions of flux / mass
ratios may improve the agreement between the theory and observations. 

As a byproduct of our simulations we have the following comment related
to the distribution of time scales. 
Many authors (\cf Wood and Mao 2005, Sumi \etal 2006,
Popowski \etal 2005) discuss distribution
of time-scales of microlensing event, providing and
comparing the values of mean time-scale in days.
It is worth to notice that both, observational histograms
(e.g. Sumi \etal 2006) as well as simulated distributions
(e.g. Paczyñski 1991, Wood and Mao 2005) are usually presented
on logarithmic abscissa. In this way the plots are better confined and 
more symmetric than their equivalent linear representation, and thus
easier to compare with a Gaussian distribution.
This is the reason we postulate using a mean logarithm
of time-scales instead of mean time-scales in any application.
Mean values derived from distributions far from Gaussian
can be mis-guiding, and in the case of microlensing, can cause
greater apparent discrepancies between models and observations, 
as small deviations at long time-scales tail give big difference in
mean value. 

It is also worth mentioning that the distribution of time-scales given
by the Galaxy model, and the distribution of the observed (or, more
precisely, fitted) values are not the same (see \eg Wyrzykowski 2005),
since longer lasting events have a greater chance to be included in the
sample of microlensing event candidates, both in real observations and
in simulations using realistic algorithm of event selection. In case of
simulations it is also possible to investigate the distribution of
intrinsic time scales in the subsample selected as microlensing event
candidates. Our calculations give the mean logarithmic Einstein
time-scale of $\approx 14^\mathrm{d}$ for all simulated events, 
$\approx 18^\mathrm{d}$ for simulated events, that were selected as
microlensing candidates, and $\approx 22^\mathrm{d}$ for time scales
fitted to the selected subsample. It is interesting that the logarithmic
means for OGLE II and OGLE III EWS events (Udalski~2003) are~$\approx 22^\mathrm{d}$ 
and~$\approx 18^\mathrm{d}$ respectively, in good
agreement with our results. 

\Acknow{We thank Andrzej Udalski and the OGLE Team for the permission of
using their unpublished data.
This work was supported in part by the Polish Ministry of Science and
Higher Education grant N203 008 32/0709.
}

\vfill
\eject
\end{document}